# A Crowd Monitoring Framework using Emotion Analysis of Social Media for Emergency Management in Mass Gatherings

**Minh Quan Ngo**
Faculty of Information Technology
Monash University
Victoria, Australia
Email: mqng01@student.monash.edu

**Pari Delir Haghighi**
Faculty of Information Technology
Monash University
Victoria, Australia
Email: pari.delirhaghighi@monash.edu

**Frada Burstein**
Faculty of Information Technology
Monash University
Victoria, Australia
Email: frada.burstein@monash.edu

## Abstract

In emergency management for mass gatherings, the knowledge about crowd types can highly assist with providing timely response and effective resource allocation. Crowd monitoring can be achieved using computer vision based approaches and sensory data analysis. The emergence of social media platforms presents an opportunity to capture valuable information about how people feel and think. However, the literature shows that there are a limited number of studies that use social media in crowd monitoring and/or incorporate a unified crowd model for consistency and interoperability. This paper presents a novel framework for crowd monitoring using social media. It includes a standard crowd model to represent different types of crowds. The proposed framework considers the effect of emotion on crowd behaviour and uses the emotion analysis of social media to identify the crowd types in an event. An experiment using historical data to validate our framework is described.

**Keywords**

Crowd modelling, crowd monitoring, social media analysis, emotion analysis, emergency management

## 1 Introduction

To maintain safety in mass gatherings, there is a need for the event organisers and emergency services to quickly detect emerging or potentially critical situations in a crowd (Wirz et al. 2012). Hence, crowd monitoring plays an important role in emergency management for large-scale events. Unfortunately, current methods used in crowd monitoring such as close-circuit television (CCTV) observation or drone surveillance are resource intensive and can be affected by human errors such as the loss of concentration (Davies et al. 1995). Recent advances in mobile computing and sensing provide an exciting opportunity to collect a wide range of new and rich data that can be used to detect emergency situations in a crowd. Examples include GPS location information in participatory sensing (Wirz et al. 2012), Bluetooth identifier collected from sensors integrated on participants' mobile phones (Weppner et al. 2013), and social media data to identify the dangerous crowd types (Delir Haghighi et al. 2013). Although each data source has its own strength and limitation when used for crowd monitoring, employing social media as the information source has certain advantages over other sources. Firstly, it is not affected by operating conditions, such as the light condition in computer vision based approaches using CCTV cameras. Secondly, it does not require deploying and operating any equipment at the venue or installing software on participants' devices, which might potentially raise privacy concerns. Despite these strengths, very limited work has been done using social media for the purpose of crowd monitoring.

Another limitation of existing studies is that they do not include a formal crowd model for uniform classification of different crowd types. The use of a unified crowd model can provide consistency and interoperability between different monitoring systems. Therefore, it is highly important to incorporate a standard crowd model into crowd monitoring approaches.

This paper proposes a novel crowd monitoring framework that performs emotion analysis of social media to identify the crowd types in a mass gathering event. The proposed framework incorporates a standard crowd model for emergency management in mass gatherings to represent different types of crowds.

This rest of this paper is structured as follows. In Section 2, we discuss the role of crowd monitoring for emergency management in mass gatherings and review the current approaches in crowd monitoring. Section 3 describes our proposed crowd monitoring framework using emotion analysis of social media. Section 4 discusses the implementation and evaluation of the proposed framework using historical data. Finally, Section 5 concludes the paper.

## 2   Crowd Monitoring In Mass Gatherings

### 2.1   Emergency Management in Mass Gatherings

A mass gathering is defined as an event attended by a large crowd of spectators and participants (Arbon 2004). According to Arbon (2007), the number of participants in a mass gathering event can be more than 25000 people. The types of an event can also vary from sporting events to concerts. Because of the large number of participants and high density as well as the impact of psychological factors such as crowd mood, there is 'the potential for a delayed response to emergencies' in mass gatherings (Arbon 2004). As a result, a great deal of attention has been given to improve the emergency management in mass gatherings, particularly in terms of planning and provision of medical and security services.

Emergency management in mass gathering events consists of three main stages: i) pre- event; ii) during the event; and iii) post-event (Delir Haghighi et al. 2013). The provision of emergency services during the event is a challenging task as it requires real-time interaction and communication between the staff as well as real-time decision-making. A review of crowd disasters at mass gathering events by Soomaroo et al. (2012) shows that there is a poor response time for emergency services in most of the cases. The knowledge about crowd behaviour during this stage can significantly improve real-time decision making and reduce emergency service response times.

According to Berlonghi (1995), safety of a mass gathering event depends on crowd management and crowd control. Crowd management includes all measures taken to facilitate the movement and enjoyment of the participants. It can be considered as a proactive effort, in contrast to the crowd control which is a reactive effort taken when the crowd is out of control or when an incident occurs. Crowd monitoring can be considered as a bridge between crowd management and crowd control because it helps to identify potential critical situations in the crowd and deliver timely medical care and response.

### 2.2   Current Approaches in Crowd Monitoring

The state-of-the-art in crowd monitoring mainly encompasses three main approaches including computer vision based methods, sensory data analysis and social media analysis. Davies et al. (1995) highlight a number of drawbacks of using CCTV systems for crowd monitoring: i) this approach is highly time consuming and labour intensive because of using a large number of cameras and generating huge data volumes, ii) human observers are likely to lose their concentration or fail to notice infrequent or gradual changes in the crowd, and iii) the cameras are not always installed in the best locations. Further, the performance of this technique can be affected by obstacles and low lighting conditions (Wirz et al. 2012).

Sensory data analysis for crowd monitoring relies on collecting data from external and internal sensors. External sensors here refer to sensors that are deployed at the venue such as sound, pressure or temperature sensors. Internal sensors are built into mobile phones like accelerometers (Roggen et al. 2011; Wirz et al. 2012). One of the benefits of using wireless sensors compared to computer vision based approaches is that they use mobile communication technologies such as Bluetooth which does not require a direct visual line of sight. However, deploying a large number of external sensors in open and large spaces can be expensive and difficult. On the other hand, collecting data from internal sensors requires installing appropriate software on the participants' mobile phones. This approach can lead to privacy issues when personal and sensitive data like GPS location is collected.

Social Network Sites (SNS) such as Twitter are considered as 'soft sensors' that enable collecting information about how people feel and think (Ramesh et al. 2014). Content analysis of data obtained from soft sensors can be used to reason about a crowd's mood and behaviour. However, there is a scarcity of research on using social media analysis for crowd monitoring. One of the existing works is introduced by Delir Haghighi et al. (2013) that uses social media analysis (sentiment analysis of tweets) to identify the crowd types in mass gatherings. However, this work does not consider the user's emotions in reasoning about crowd types.

The literature also shows that most of crowd monitoring approaches are not underpinned by a standard crowd modelling approach. The next subsection discusses crowd modelling in mass gatherings.

## 2.3 Crowd Modelling

Using a formal crowd model in crowd monitoring approaches can provide consistency in data integration and across three different phases of mass gathering management (Delir Haghighi et al. 2013). However, there have been a very limited number of studies on crowd modelling. Berlonghi (1995) suggested that the distinction between different crowd types using a crowd model is crucial in crowd control and crowd management. His proposed crowd model is widely adopted in emergency management literature (EMA 1999; FEMA 2005). The model consists of eleven crowd types: *ambulatory crowd*, *disability/limited movement crowd*, *cohesive/spectator crowd*, *expressive/revellous crowd*, *participatory crowd*, *aggressive/hostile crowd*, *demonstrator crowd*, *escaping/trampling crowd*, *dense/suffocating crowd*, *rushing/looting crowd* and *violent crowd*.

Each crowd type is described according to the purpose of the gathering and the activities in the crowd. However, using only the definitions, it is very difficult for a computer-based approach to distinguish between different crowd types because the model does not provide any distinct or measurable features/attributes.

Another notable work on the classification of crowd is from Lofland (1985) who categorised a crowd by the motivating emotions: *anger*, *fear* and *joy*. This classification was based on the studies on collective behaviour and emotions (Brown 1954; Lofland 1985; Smelser 1998). These studies suggest that emotions can be used as a feature to distinguish different crowds. Yet, this approach will require a mapping between a crowd type and its associated emotions.

The following section describes our proposed crowd monitoring framework that aims to address the main limitations of current works. It provides emotion analysis of social media data and a mapping model to further improve reasoning about different crowd types. It also includes a formal and well-adopted crowd model.

## 3 A Crowd Monitoring Framework Using Emotion Analysis of Social Media

This paper proposes a novel framework for crowd monitoring in mass gatherings that incorporates emotion analysis of social media as shown in Figure 1. Social media is firstly probed to get the context data about users (i.e. the crowd in mass gatherings). Then emotion analysis is performed to extract high-level context data that is the emotional states of the crowd. Finally, these emotional states are utilized as the attributes or features to map a crowd into specific crowd types.

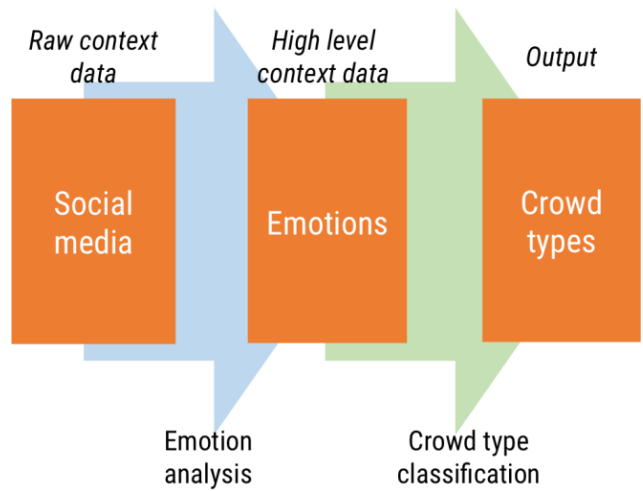

*Figure 1: The process to identify a crowd type from social media*

Figure 2 shows the main components of our framework that include: i) Context data; ii) Emotion analysis; iii) Emotion model; iv) Crowd model; v) Emotion - Crowd type mapping model; vi) Rule based reasoning. The following subsections describe each component.

### 3.1　Context Data

The first component of the framework is the *Context Data* component. This research focuses on the use of social media, or Twitter in particular, as the information source for context data. Twitter has a large volume of users and its public APIs make the data highly accessible. The user-generated content is mostly text-based and in the form of a short message which has no more than 140 characters, known as a tweet. A tweet might also contain information beyond the text. A significant number of tweets are geo-tagged, which means they are associated with the location information. Another useful information that might exist in a tweet is hashtags. Hashtags are created by the users, and usually are included in a tweet to mark a topic of interest (Mohammad & Kiritchenko, 2015). Geolocation data and hashtags enable filtering the tweets based on a particular event.

Using Twitter as the information source for crowd monitoring, contextual data can be also collected in real-time. This enables real-time monitoring which is essential in mass gatherings where the dynamic nature of a crowd can suddenly change from a calm type to an aggressive type (Berlonghi 1995).

### 3.2　Emotion Model

In our framework, an emotional factor is emphasised as the common motivation of the behaviour in a crowd. Different emotions can consequently lead to different crowd behaviours. For example, according to Lofland's typology of spontaneous collective behaviours (Kornblum 2011), panic exodus from burning theatre is motivated by *fear* whereas a race riot is caused by *anger*. Therefore, an *emotion model* is essential to distinguish between different emotions of the people in a crowd.

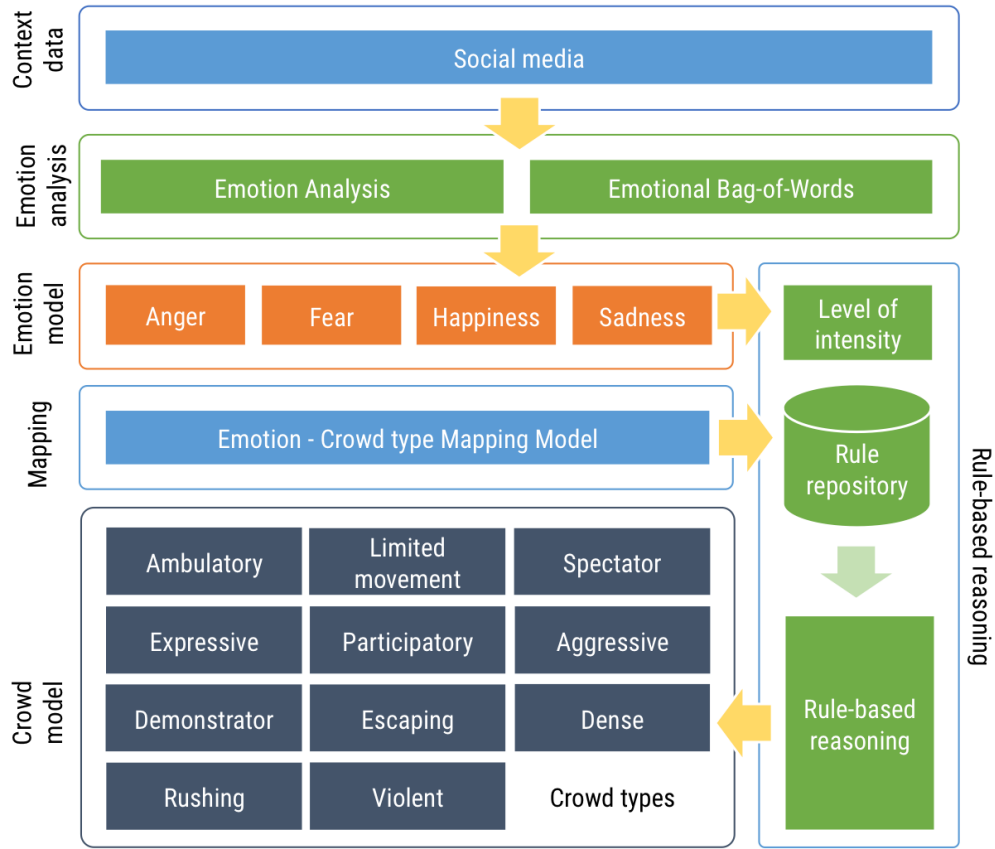

*Figure 2: An overview of the crowd monitoring framework using emotion analysis of social media*

While it is often assumed that among different theories of emotions there exists a small set of basic or fundamental emotions, there is little agreement on determining those basic emotions (Ortony et al. 1990). Based upon the theory of evolution of the biological process, Plutchik (2001) identified eight basic emotions consisting of four pairs of opposite emotions: *anger/fear*, *joy/sadness*, *trust/disgust*, *anticipation/surprise*. Investigating the facial expression of humans, Ekman et al. (1971) suggested there are six basic emotions that can be universally recognised, regardless of the language. The six basic emotions are *anger*, *fear*, *happiness*, *sadness*, *surprise* and *disgust*. Ekman's model has been adopted in many studies related to human emotions (Alm et al. 2005; Mohammad et al. 2015; Roberts et al. 2012). A recent research by Jack et al. (2014) suggest that there is a similarity between *anger* and *disgust*, and also between *surprise* and *fear* in terms of early facial expressions. For example, both *anger* and *disgust* share a wrinkled nose, while both *surprise* and *fear* share raised eyebrows.

Considering different classifications of emotions, we adopt four major and distinct basic emotions of *anger*, *fear*, *happiness* and *sadness* as our emotion model in the proposed framework.

### 3.3  Emotion Analysis

The *Emotion Analysis* component uses a Bag-of-Words approach illustrated in Figure 3. The Bag-of-Words method is based on a simplified representation of a document introduced by Joachims (1996). This representation only considers the frequency of appearance of the words in a document regardless of the order of the words. In our approach, a document is a single tweet and a word is a uni-gram that builds up the tweet. The process that splits a document into a sequence of words is called tokenizer.

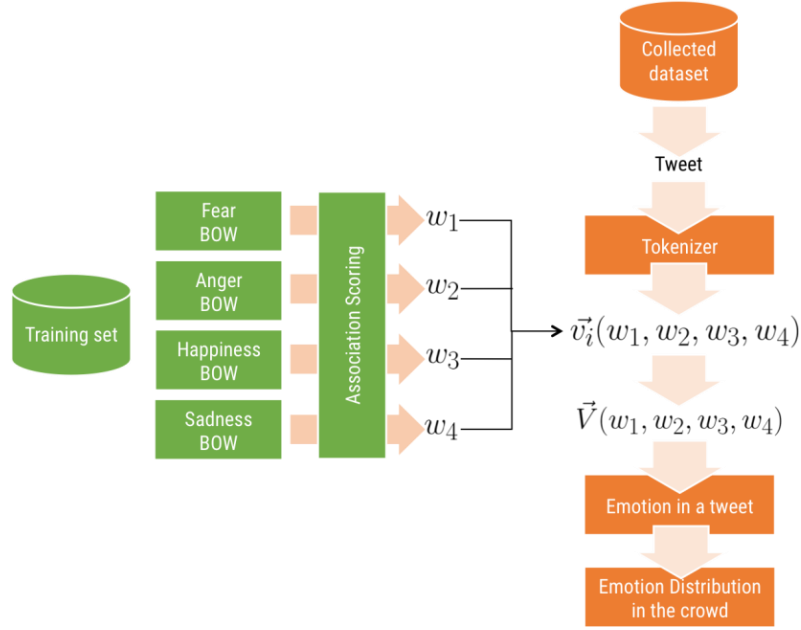

*Figure 3: An overview of the Bag-of-Words approach*

The training set, or labelled corpus, is a collection of tweets labelled with actual emotions in order to serve as the training data for the classifiers. In this paper, the NRC Hashtag Emotion Corpus (Mohammad et al. 2015) was selected as the labelled corpus. This NRC Hashtag Emotion Corpus was annotated with Ekman's six basic emotions, which contain our chosen four emotions: *anger, fear, happiness and sadness*.

In the Association Scoring process, the association of a word with an emotion: *anger*, *fear*, *happiness* or *sadness* is measured by a normalised weight. This weight represents the correlation of the appearance of the word in a tweet with the emotion extracted from that tweet. The reason for normalisation is that a weight must be comparable with the weights of other words and other emotions. We adopted the SoA score proposed in the NRC Hashtag Emotion Lexicon by Mohammad et al. (2015) for this purpose. In their paper, the SoA of a word toward an emotion is calculated from the frequency of this word appearing in the tweets tagged with that emotion. Using the SoA score, the emotional status of a word can be modelled by a four-dimensional vector, in which each dimension represents an emotion of the basic emotions: *anger, fear, happiness* and *sadness*. This vector is proposed as the emotional weight vector as in the formula below.

$$\vec{v}_i(w_1, w_2, w_3, w_4)$$

where $\vec{v}_i$ is an emotional weight vector of a word and $w_1, w_2, w_3, w_4$ are four weights that represent its association with the four emotions: *anger*, *fear*, *happiness* and *sadness* respectively. It can be noted that because the SoA score only represents the association of a word toward an emotion, the normalised weight only has a positive value.

A tweet can be considered as a set of words, each of which contributes to the overall emotional status of the tweet. Therefore, the emotional status of a tweet can also be represented by a four-dimensional vector that is the summative of the emotional weight vectors of the elemental words.

$$\vec{V} = \sum_{i=1}^{n} \vec{v}_i(w_1, w_2, w_3, w_4)$$

where $\vec{V}$ is the emotional weight vector of the tweet, $\vec{v}_i$ is a weight vector of word *i* and *n* is the total number of words in the tweet. The summative vector $\vec{V}$ illustrates how close the whole tweet is from each emotion. The dominant emotion of the tweet is selected as the dimension in $\vec{V}$ that has the highest magnitude. The tweet will then be labelled with that dominating emotion.

The dominating emotion, in fact, represents the emotion of an individual in the crowd rather than the crowd as a whole. The relationship between an individual's emotion and a group's emotion has been discussed in several studies. Barsäde et al. (1998) suggest that while the group emotion can be felt by each individual member, the group emotion can also be built from the individuals' emotions in a bottom-up approach. This theory confirms that the crowd emotion can be determined from the emotion of each individual.

Since the individuals in a crowd might experience different emotional states, a crowd can be associated with multiple emotions. This can be represented by the distribution of emotion in the crowd. In our study, the chosen measure to represent the distribution is the *emotion rate*. The emotion rate in a crowd in a given time period can be calculated by the following formula.

$$rate(e_i) = \frac{n(e_i)}{\sum_{i=1}^{4} n(e_i)}$$

where $rate(e_i)$ is the emotion rate of the emotion $e_i$ and $n(e_i)$ is the number of tweets classified with emotion $e_i$. The emotion rate $rate(e_i)$ can have any value from 0 to 1 ($0 \leq rate(e_i) \leq 1$), indicating how strong the emotion $e_i$ is during the sampling period. In order to detect the change of the emotional state in the crowd, the calculation of emotion rate is performed repetitively after an interval *t*.

In the proposed framework, emotion rates are used by the mapping model to reason about a crowd type. Thus, it is important that the strength of the emotions are also considered during the mapping. The stronger the emotion rate, the higher is the confidence in the occurrence of a crowd type. The level of an emotion can be labelled as *low* or *high* by comparing the emotion rate of an emotion against a threshold. We use a threshold *th* to distinguish between the low and high levels of the emotions.

Assigning the threshold value is application specific and depends on several factors. People tend to post tweets when they feel a certain type of emotion, and therefore thresholds can vary for different emotions. The study by (Mohammad et al. 2015) shows that the percentage of joyful tweets is much higher than the other emotions. Because of the cultural or regional differences, the emotional content shared on Twitter might be also different across countries (Larsen et al. 2015). Even among the same group of users, emotional content can vary during different time periods. Analysing 509 million tweets, Golder et al. (2011) report that people are generally less positive during winter than summer.

In our experiment, we use a simple approach to determine the threshold based on the moving average and z-score of the emotion rate using an experimental dataset collected about a specific mass gathering event.

### 3.4 Emotion - Crowd Type Mapping Model

The crowd monitoring framework also includes two components of the *Crowd Model* and *Emotion-Crowd Mapping Model*. To represent different crowd types, the framework incorporates a well-adopted and widely used crowd model introduced by Berlonghi (1995) which consists of eleven crowd types (discussed in Section 2.3). However, the model provides only descriptions of each crowd type that are insufficient to automatically identify the type of a certain crowd. Therefore, there is a need to introduce attributes or features to the existing model to enable distinguishing between crowd types. We extend Berlonghi's model by introducing four basic emotions as additional features/attributes and use them for mapping emotions to crowd types through an Emotion - Crowd type Mapping Model. The basic emotions include *anger*, *fear*, *happiness* and *sadness* denoted as $e_1, e_2, e_3, e_4$ respectively.

Based on a wide range of studies on social behaviour and psychology that explore the relationship between emotions and crowd types (Kornblum 2011; Ray 2014; Zeitz et al. 2009; Ziemann et al. 2009), we identify five distinct groups of crowd types according to their common motivating emotions. Table 1 shows these five groups and their associated crowd types and emotions.

| Group | Crowd types | Motivating emotion |
|---|---|---|
| Group 1 | Ambulatory $c_1$, Limited movement $c_2$, Spectator $c_3$ | None |
| Group 2 | Expressive/Cohesive $c_4$, Participatory crowd $c_5$ | Happiness $e_3$ |
| Group 3 | Aggressive crowd $c_6$, Demonstrator $c_7$, Violent crowd $c_{11}$ | Anger $e_1$ |
| Group 4 | Escaping crowd $c_8$, Dense/Suffocating crowd $c_9$ | Fear $e_2$ |
| Group 5 | Rushing/Looting crowd $c_{10}$ | Anger $e_1$, sadness $e_4$ |

*Table 1: Groups of crowd types based on common motivating emotion*

**Group 1 -** This group is not associated with any emotion. From the emergency management point of view, the three crowd types in this group do not pose any potential danger. If the accelerometer data can be collected from the built-in sensors in mobile phones of the participants, it can be used to recognise user activities (e.g. sitting, walking, running, etc.). This knowledge can be used as an additional crowd feature for making distinction between these crowds. For example, an ambulatory crowd involves mainly walking, while a crowd of spectators or limited movement can be identified by sitting or standing activities.

**Group 2** – This group is characterised by happiness. An expressive crowd can be associated with emotional release such as cheering or singing. A participatory crowd is the crowd involved in the actual activities of an event. Thus, it is very difficult to distinguish between these two types based on their physical activities.

**Group 3** – The third group is associated with anger. Compared to a demonstrator crowd, a violent crowd involves activities that have a higher level of motion such as running and walking.

**Group 4** – This group is motivated by fear. A dense/suffocating crowd is characterised by high density and lack of movement, while an escaping crowd can be associated with the walking and running activities.

**Group 5** – The fifth group includes only one crowd type that is the rushing/looting crowd but is associated with two dominant emotions of anger and sadness. Therefore, this crowd type can be easier to identify.

### 3.5  Rule Based Reasoning

The Rule Based Reasoning component is responsible for applying the rules defined in the Rule Repository. The input of this component is the level of emotions obtained from the Emotion Analysis. The distinction between different levels of the emotions (i.e. low and high) is determined using a threshold value as discussed in Section 3.3 and based on the following formula:

$$rate(e_i) > th(e_i) \Rightarrow level(e_i) = high$$

The output is a crowd type $c_i$ that best describes the state of the crowd in the event. It is notable that each rule is processed independently and this allows multiple crowd types to exist in an event. According to the groups of crowd types and their associated motivating emotions shown in Table 1, the following rules are defined:

Rule 1: $(level(e_1) \neq high) \wedge (level(e_2) \neq high) \wedge (level(e_3) \neq high) \wedge (level(e_4) \neq high) \Rightarrow (c = c_1) \vee (c = c_2) \vee (c = c_3)$

Rule 2: $(level(e_3) = high) \Rightarrow (c = c_4) \vee (c = c_5)$

Rule 3:   $(level(e_1) = high) \Rightarrow (c = c_6) \vee (c = c_7) \vee (c = c_{11})$

Rule 4:   $(level(e_2) = high) \Rightarrow (c = c_8) \vee (c = c_9)$

Rule 5:   $(level(e_1) = high) \wedge (level(e_4) = high) \Rightarrow (c = c_{10})$

## 4  Implementation and Evaluation

We implemented the proposed crowd monitoring framework as a simple Java application. In order to evaluate the framework, we conducted an evaluation using real data about a past sporting event where the crowd types were already known. Our experiment involved data collection and content analysis of tweets using our proposed Emotion - Crowd type Mapping Model and Rule Based Reasoning. The results were then compared to the identified crowd types to verify the accuracy of our approach.

### 4.1  Data Collection

Since the Bag-of-Words were constructed from an English corpus, we limited our search to the events that occurred in an English speaking country. Further, we filtered our search to the sporting events because these events tend to involve a large number of participants experiencing strong (i.e. high level) emotions.

We selected a boxing match in USA between Floyd Mayweather and Marcos Maidana on 3th May 2014 where a stampede occurred after the fight. The boxing match was held at the Grand Garden Arena of the MGM Grand Hotel in Las Vegas, USA with more than 16,000 people attended. The match started at 6:00PM local time (UTC-7). The stampede occurred around 10:45PM during the post-fight media conference. The police received an emergency call at 10:45PM reporting a gunshot at the venue. According to the official statement, when the fans were leaving the arena, the stampede was triggered by a loud bang. The noise was caused by a falling partition but it was mistaken for a gunshot. The incident caused a massive panic and people started rushing to the exits. More than 50 people were trampled and crushed in the stampede and suffered from minor injuries. The situation got under control by 1:00AM the next day. The layout of the arena was later criticized for having only two exits and the pathway was too narrow which caused a bottleneck. Since it was confirmed that a stampede crowd had occurred in the event, the crowd types for this event could be identified as *escaping* and *dense/suffocating* crowds. According to our classification presented in Table 1, these two crowd types belong to Group 4 motivated by *fear*.

We collected 33,935 tweets from the Twitter Advanced Search[1]. Since the venue of the boxing match was MGM Grand Hotel, we used the hashtag #MGMGrand. We also limited our search to the tweets posted over the course of one week (between dates: 2014-05-01 to 2014-05-07) in order to consider the crowd emotions before and after the stampede.

After data collection, emotion analysis was performed to extract the emotions from the tweets using the Bag-Of-Words approach and NRC Hashtag Emotion Lexicon (Mohammad et al. 2015). The experiment was performed in a manner that it simulated the real-time analysis. The dataset was divided into smaller segments $t_i$ and the emotion distribution in each $t_i$ was calculated. In the experiment, interval *t* for a segment was set as 15 minutes.

The next step was to determine the thresholds and to label the levels of the emotions for each segment. As mentioned in Section 3.3, the values of thresholds are application-specific and can be different for each emotion. The rates of *anger*, *fear*, *happiness* and *sadness* over time were normally distributed in our dataset, therefore, a method to calculate the thresholds was implemented using the moving average and z-score for each segment $t_i$. The threshold was chosen as 1.0. Finally, the Rule Based Reasoning was implemented with a set of rules as defined in Section 3.5. The crowd groups (and crowd types) that occurred in each $t_i$ were inferred based on the levels of the four emotions. The analysis results were compared with the known crowd types (i.e. escaping and dense/suffocating crowds). In term of timeliness, our inference was evaluated against the time that the accident was reported to police according to the official announcement.

---

[1] https://twitter.com/search-advanced

## 4.2　Evaluation

We performed the Emotion Analysis using NRC Hashtag Emotion Lexicon (Mohammad et al. 2015) on the dataset. As a result, 28,698 tweets were labelled with one of four emotions: *anger*, *fear*, *happiness* and *sadness*. Overall, the Bag-of-Words approach successfully extracted the emotion from 84.67% of the collected tweets. Our statistical analysis was performed on the labelled data. The boxing match started at 6:00PM local time on May 3rd. While the stampede occurred around 10:45PM, it was reported that the situation got under control at 1:00AM. This information was helpful to identify the time period that could be related to the event. The event was defined as a seven-hour period starting from 6:00PM to 1:00 AM. Our statistical analysis concentrated on this period. *Figure 4* shows the distribution of emotions during this event. As can be seen, the dominating emotion was anger. The reason might be that due to the nature of the boxing match the tweets contained a large number of aggressive words.

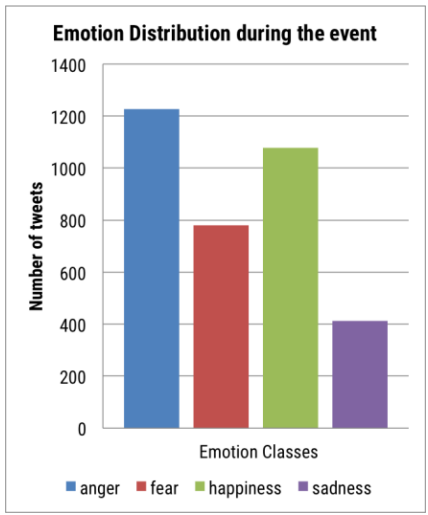

*Figure 4: The number of tweets labelled with each emotion in the dataset during the event*

Figure 5 illustrates the emotion rates of *anger*, *fear*, *happiness* and *sadness* during the event. In order to detect the abnormal situation from the emotion rates over time, we applied a combined method of moving average and z-score to identify the sudden changes that significantly deviated from the expected values.

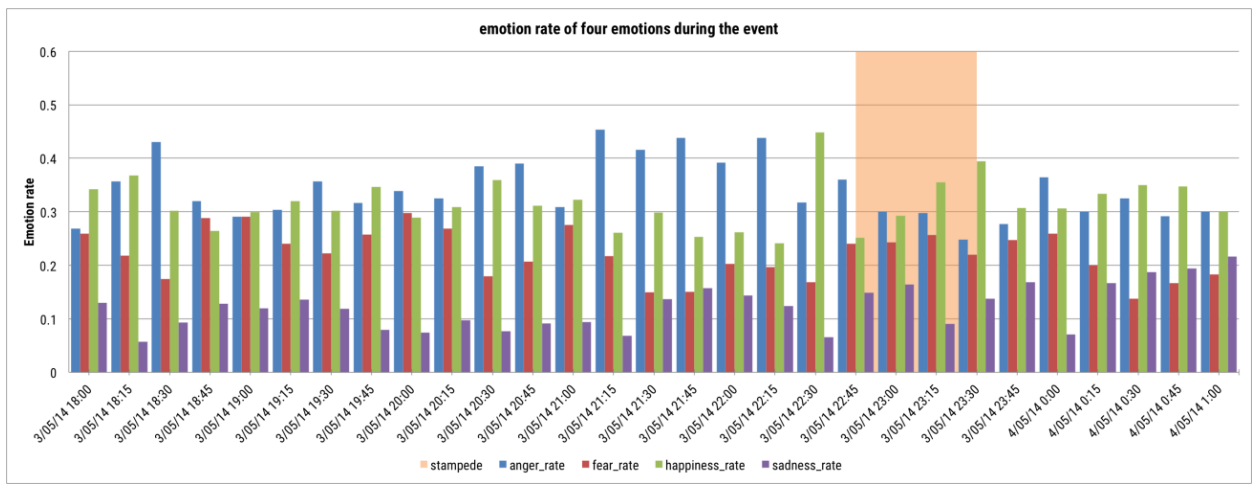

*Figure 5: The percentage of anger, fear, happiness and sadness over time during the event*

Figure 6 focuses on the one-hour period when the stampede happened. The threshold line represents our chosen high-level threshold of 1.0. It was noticeable that among four emotions, only *fear* exceeded this threshold in two segments at 10:45PM and 11:00PM. It showed that when the stampede occurred there was a significant increase in the proportion of tweets labelled with *fear*, suggesting a high level of *fear* in the crowd. Applying the Rule Based Reasoning, the crowd type was inferred to be under Group 4, which included the escaping crowd and the dense/suffocating crowd.

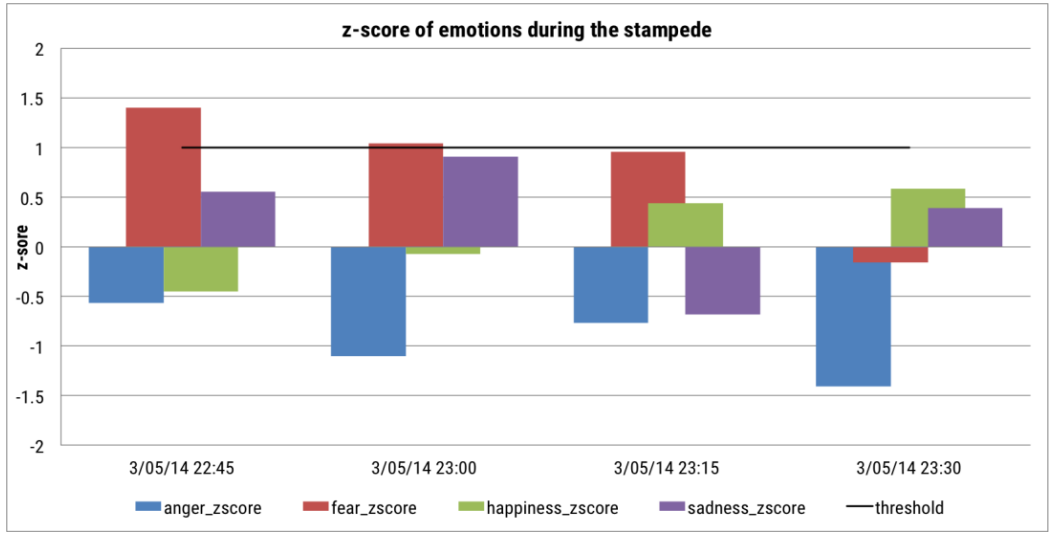

*Figure 6: Threshold and z-score of each emotion during the stampede*

In terms of accuracy, the inferred crowd types matched with the reported crowd types (i.e. escaping and dense/suffocating crowds). Although our defined rules only identify the group of crowd types rather than a certain crowd type, in this case study, the rules were sufficient to identify a dangerous situation. Regarding timeliness, using the proposed emotion analysis method, the escaping and dense/suffocating crowds were detected in the 10:45PM segment by an increase in the emotion level of *fear* as shown in Figure 6. This was the time when the local police received the emergency call about the stampede. In real-time monitoring, a shorter time interval can be used to improve the timeliness of access to the analysis results.

## 5　Conclusion

This paper introduced a novel framework for crowd monitoring in mass gatherings that applies emotion analysis of social media data. The framework incorporated a well-adopted crowd model introduced by Berlonghi (1995). Realising the impact of emotions on the behaviour of the crowd (Kornblum 2011), our research considered four emotional states as additional features to classify different types of crowds. These emotions included: *anger*, *fear*, *happiness* and *sadness* adopted from (Ekman et al. 1971). Based on the studies on social behaviour and psychology, an Emotion - Crowd Type Mapping Model was also proposed to map emotions to the crowd types.

An experiment using real data about a past boxing match where a stampede occurred was conducted. The results showed that the framework was able to detect the correct crowd types in a timely manner.

There are several aspects of our proposed framework that can be further improved such as providing support for real-time data collection and reasoning, and integrating mobile sensing particularly activity recognition into the framework to improve accuracy of distinguishing between different crowd types.

# 6   References

Alm, C. O., Roth, D., and Sproat, R. 2005. "Emotions from text: machine learning for text-based emotion prediction," in *Proceedings of the co nference on Human Language Technology and Empirical Methods in Natural Language Processing*, pp. 579-586.

Arbon, P. 2004. "The development of conceptual models for mass-gathering health," *Prehospital and Disaster Medicine* (19:03), pp 208-212.

Arbon, P. 2007. "Mass-gathering medicine: a review of the evidence and future directions for research," *Prehospital and disaster medicine* (22:02), pp 131-135.

Barsäde, S. G., and Gibson, D. E. 1998. "Group emotion: A view from top and bottom," *Research on managing groups and teams* (1), pp 81-102.

Berlonghi, A. E. 1995. "Understanding and planning for different spectator crowds," *Safety Science* (18:4), pp 239-247.

Brown, R. W. 1954. "Mass phenomena," *Handbook of social psychology* (2), pp 833-876.

Davies, A. C., Yin, J. H., and Velastin, S. A. 1995. "Crowd monitoring using image processing," *Electronics & Communication Engineering Journal* (7:1), pp 37-47.

Delir Haghighi, P., Burstein, F., Li, H., and Wang, C. 2013. "Integrating social media with ontologies for real-time crowd monitoring and decision support in mass gatherings," in *PACIS 2013 Proceedings*, pp. 1-14.

Ekman, P., and Friesen, W. V. 1971. "Constants across cultures in the face and emotion.," *Journal of personality and social psychology* (17:2), p 124.

EMA, A. E. M. 1999. "Manual 2 - Safe and Healthy Mass Gatherings," *A Health, Medical and Safety Planning Manual for Public Events. Commonwealth of Australia: Dickson ACT*).

FEMA 2005. "Special Events Contingency Planning Manual,").

Golder, S. A., and Macy, M. W. 2011. "Diurnal and Seasonal Mood Vary with Work, Sleep, and Daylength Across Diverse Cultures," *Science* (333:6051), pp 1878-1881.

Jack, R. E., Garrod, O. G., and Schyns, P. G. 2014. "Dynamic facial expressions of emotion transmit an evolving hierarchy of signals over time," *Current biology* (24:2), pp 187-192.

Joachims, T. 1996. "A Probabilistic Analysis of the Rocchio Algorithm with TFIDF for Text Categorization.," DTIC Document: Carnegie-Mellon Univ Pittsburgh, Dept Of Computer Science.

Kornblum, W. 2011. *Sociology in a changing world*, (Cengage Learning.

Larsen, M., Boonstra, T., Batterham, P., O'Dea, B., Paris, C., and Christensen, H. 2015. "We Feel: Mapping emotion on Twitter," *Biomedical and Health Informatics, IEEE Journal of* (PP:99), pp 1-1.

Lofland, J. 1985. *Protest: Studies of collective behavior and social movements*, (Transaction Publishers.

Mohammad, S. M., and Kiritchenko, S. 2015. "Using hashtags to capture fine emotion categories from tweets," *Computational Intelligence* (31:2), pp 301-326.

Ortony, A., and Turner, T. J. 1990. "What's basic about basic emotions?," *Psychological review* (97:3), pp 315-331.

Plutchik, R. 2001. "Integration, Differentiation, and Derivatives of Emotion," *Evolution and Cognition* (7:2), pp 114-125.

Ramesh, M. V., Shanmughan, A., and Prabha, R. 2014. "Context aware ad hoc network for mitigation of crowd disasters," *Ad Hoc Networks* (18), pp 55-70.

Ray, L. 2014. "Shame and the city-'looting', emotions and social structure," *The Sociological Review* (62:1), pp 117-136.

Roberts, K., Roach, M. A., Johnson, J., Guthrie, J., and Harabagiu, S. M. 2012. "EmpaTweet: Annotating and Detecting Emotions on Twitter.," in *Proceedings of the Eight International Conference on Language Resources and Evaluation (LREC'12). European Language Resources Association (ELRA)*: Istanbul, Turkey, pp. 3806-3813.

Roggen, D., Wirz, M., Tröster, G., and Helbing, D. 2011. "Recognition of Crowd Behavior from Mobile Sensors with Pattern Analysis and Graph Clustering Methods," in *arXiv preprint arXiv:1109.1664*.

Smelser, N. J. 1998. *The social edges of psychoanalysis*, (Univ of California Press.

Soomaroo, L., and Murray, V. 2012. "Disasters at mass gatherings: lessons from history," in PLoS currents.online: Published online 2012 March 12.doi:10.1371/currents.RRN1301

Weppner, J., and Lukowicz, P. 2013. "Bluetooth based collaborative crowd density estimation with mobile phones," in *Pervasive Computing and Communications (PerCom), 2013 IEEE International Conference on*, IEEE, pp. 193-200.

Wirz, M., Franke, T., Roggen, D., Mitleton-Kelly, E., Lukowicz, P., and Troster, G. 2012. "Inferring crowd conditions from pedestrians' location traces for real-time crowd monitoring during city-scale mass gatherings," in *Enabling Technologies: Infrastructure for Collaborative Enterprises (WETICE), 2012 IEEE 21st International Workshop on*, IEEE, pp. 367-372.

Zeitz, K. M., Tan, H. M., Grief, M., Couns, P., and Zeitz, C. J. 2009. "Crowd behavior at mass gatherings: a literature review," *Prehospital and disaster medicine* (24:01), pp 32-38.

Ziemann, A. E., Allen, J. E., Dahdaleh, N. S., Drebot, I. I., Coryell, M. W., Wunsch, A. M., Lynch, C. M., Faraci, F. M., Howard, M. A., Welsh, M. J., and others 2009. "The amygdala is a chemosensor that detects carbon dioxide and acidosis to elicit fear behavior," *Cell* (139:5), pp 1012-1021.

## Acknowledgements

This research is partly supported by the Federation University 'Self-sustaining Regions Research and Innovation Initiative', an Australian Government Collaborative Research Network (CRN) grant.